\documentclass[preprint]{aastex}
\slugcomment{}

\shorttitle{On the Origin of HD149026b} 
\shortauthors{Ikoma et al.}

\begin{document}

\newcommand{\mearth}{{\rm M}_\oplus}
\newcommand{\sub}[1]{_{\rm #1}}

\title{On the Origin of HD149026b}

\author{M. Ikoma\altaffilmark{}}
\affil{Research Center for the Evolving Earth and Planets, 
Tokyo Institute of Technology, Ookayama, Meguro-ku, Tokyo 152-8551, 
Japan}
\email{mikoma@geo.titech.ac.jp}

\author{T. Guillot\altaffilmark{}}
\affil{Observatoire de la C\^{o}te d'Azur, 
CNRS UMR 6202, BP 4229, 06304 Nice Cedex 4, France}

\author{H. Genda, T. Tanigawa, \& S. Ida\altaffilmark{}}
\affil{Earth and Planetary Sciences, 
Tokyo Institute of Technology, Ookayama, Meguro-ku, Tokyo 152-8551, 
Japan}

\begin{abstract}

The high density of the recently discovered close-in extrasolar planet
HD149026b suggests the presence of a huge core in the planet, 
which challenges planet formation theory. 
We first derive constraints on the amount of heavy
elements and hydrogen/helium present in the planet: We find that
preferred values of the core mass are between 50 and $80\mearth$,
although a minimum value of the core mass is $\sim 35\mearth$ in the
extreme case of formation of the planet at $> 0.5\,$~AU, followed by
late inward migration after $>$ 1\,Ga and negligible reheating due
to tidal dissipation.
We then investigate the possibility of subcritical core accretion as
envisioned for Uranus and Neptune.  We show that a massive core 
surrounded by an envelope in hydrostatic equilibrium 
with the gaseous disk may indeed grows beyond
$30\mearth$ provided the core accretion rate remains larger than $\sim
2\times 10^{-5}\mearth\,\rm yr^{-1}$. 
However, we find the subcritical accretion scenario is very
unlikely in the case of HD149026b for at least two reasons: (i)
Subcritical planets are such that the ratio of their core mass to
their total mass is above $\sim 0.7$, in contradiction with
constraints for all but the most extreme interior models of HD149026b;
(ii) High accretion rates and large isolation mass
required for the formation of a subcritical $>35\mearth$ core are
possible only at specific orbital distances in a disk with a surface
density of dust equal to at least 10 times that of the minimum mass
solar nebula. This value climbs to 30 when considering a $50\mearth$
core. 
These facts point toward two main routes for the formation of this
planet: (i) Gas accretion that is limited by a slow viscous inflow of
gas in an evaporating disk; (ii) A significant modification of the
composition of the planet after gas accretion has stopped.  These two
routes are not mutually exclusive.  Illustrating the second route,
we show that for a wide range of impact parameters, giant impacts lead
to a loss of the gas component of the planet and thus may lead to
planets that are highly enriched in heavy elements. Alternatively, the
planet may be supplied with heavy elements by planetesimals 
by secular perturbations.
Both in the giant impact and the secular perturbation scenarios, we
expect an outer giant planet to be present. Observational studies by
imaging, astrometry and long term interferometry of this system are
needed to better narrow down the ensemble of possibilities.
\end{abstract}


\section{INTRODUCTION \label{sec:introduction}}

The combined detection of extrasolar planets by radial velocimetry and
transit photometry provides unique information by the measurement of
both their mass and radius. Further spectroscopic studies of the
system in and off transit can then even inform us on 
chemical species in the planet's atmosphere, its possible
evaporation, global atmospheric temperature, possible presence of
winds, etc.

Nine transiting extrasolar planets have been discovered so far 
(see www.obspm.fr/planets). 
Eight of them have radii of 1.0--1.3 Jupiter's radius (${\rm R}\sub{J}$), 
but one stands out: HD149026b has a radius of
only $0.73\pm 0.03 \rm R_{J}$  
for a mass of $0.36 \pm 0.03$ Jupiter's mass 
($= 110 \pm 10 \mearth$)
and an orbital distance of
0.042\,AU \citep{Sato05,Charbonneau06}. The parent star is believed to
be 1.3\,M$_\odot$, and is metal-rich with $\rm [Fe/H]=0.36\pm
0.05$. Contrary to the other transiting extrasolar planets, believed
to be mainly formed with hydrogen and helium
\citep{Burrows00,Bodenheimer01,GuillotShowman02,Laughlin05,Guillot05,Baraffe06}, 
HD149026b is {\it clearly} made of a significant fraction of heavy
elements. 

Indeed, based on evolution models, \citet{Sato05} and then
\citet{Fortney06} have shown the planet has a huge rock/ice core of
$\sim 70\mearth$. As noted by \citet{Sato05},
such a big core points toward formation of the planet in the core
accretion scenario \citep*[e.g.,][]{Mizuno80, BP86}, rather than
in the disk instability scenario \citep[e.g.,][]{Boss97}. 
However, HD149026b may even challenge the core accretion scenario, 
because the core mass is far larger than that of Jupiter or Saturn, and
the ``canonical'' critical core mass as derived by
\citet{Mizuno80} is around $10\mearth$. 

The purpose of this paper is to constrain the structure of HD149026b,
to examine how the planet may have formed based on these constraints,
and, hopefully, to stimulate further studies of the formation of
close-in giant planets.  In section~\ref{sec:present_structure}, we
attempt to constrain the structure of HD149026b based on various
evolution models. In section~\ref{sec:accretion}, 
we then examine whether these constraints are
compatible with the properties of growing protoplanetary cores
embedded in a gaseous circumstellar disk. On these bases, various
possibilities for the formation of this planet are explored in section
\ref{sec:possibilities}. Section \ref{sec:summary} summarizes the
results and proposes two possible formation scenarios that account for
the properties of HD149026b.

\section{PRESENT STRUCTURE: HOW MUCH HEAVY ELEMENTS? \label{sec:present_structure}}

As already evident from the measurements and from previous models of
the evolution of the planet \citep{Sato05,Fortney06}, 
HD149026b is surprisingly small, and should possess either a big core
or a large amount of heavy elements in its interior. 
In this work, we
attempt to constrain the core mass by using a variety of evolutionary
models beyond those explored by Sato et al. and Fortney et al. 
In doing so, we will show that HD149026b indeed contains a
significant amount of heavy elements, in apparent contradiction with
current models of the formation of giant planets.

\subsection{Standard Models \label{sec:standard_model}}

The evolution of a giant planet is essentially governed by a 
Kelvin-Helmholtz cooling and contraction similar to a stellar pre-main
sequence evolution but slightly modified by degeneracy effects, and by
the intense stellar irradiation that slows the cooling through the
growth of a deep radiative region that can extend to
kbar levels \citep{Guillot96, Guillot05}. 

The level of the irradiation received by the planet is calculated from
the observations by \citet{Sato05}: We adopt values of the stellar
radius ($R_*=1.45\,\rm R_\odot$) and temperature ($T_*=6150\,$K) and
the orbital distance of $0.046\,$~AU to derive zero-albedo equilibrium
temperature $T_{\rm eq}^*=1740\,$K. We use this value as a minimum for
the atmospheric temperature at the 1 bar level (the corresponding
temperature found by \citet{Fortney06} is between 2000 and 2200\,K,
and using the model of \citet{IBG05}, we find a value of 1980\,K). We
use the 1 bar level as outer boundary condition for the evolution
models \citep[see][]{Guillot05}.  

Standard models are calculated on the basis of structure consisting
of a central rock/ice core and of an envelope of solar composition,
and further assuming that the planet has either
formed {\it in situ} or moved rapidly to its present location, so that
throughout its evolution it has received a constant stellar heat flux
to slow its cooling. 

For the core, we use the equations
of state for ``rocks'' and ``ices'' obtained by \citet{HM89}. 
For the envelope, we use the H-He EOS from \citet*{SCVH95}.
The structure of the core with \citet{HM89}'s EOS 
is independent of its temperature, but this
simplification is reasonable because it has much smaller effects
than changes in the (unknown) core composition. Although it has little
effect on the models, we assume that the core luminosity is due both
to radioactive decay (with $l_{\rm chondritic}\approx
10^{20}\rm\,erg\,s^{-1}\,\mearth^{-1}$)  and cooling (assuming that in the
core the temperature is uniform and the specific heat is $c_v\approx
10^7\rm\,erg\,g^{-1}\,K^{-1}$). 
Internal opacities are
calculated from Rosseland opacity tables provided by
\citet{Allard01}.

\begin{figure}
\plotone{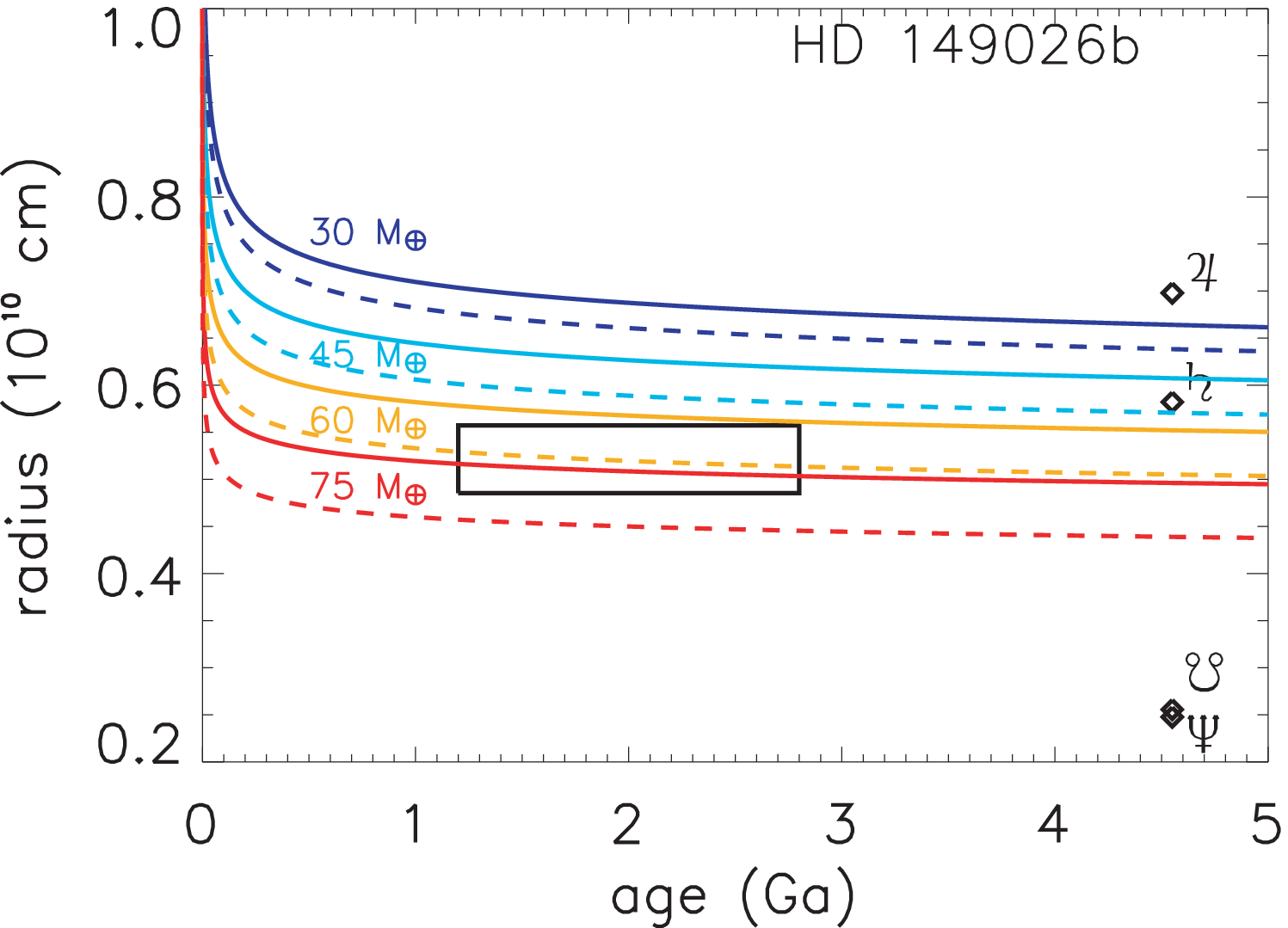}
\caption{Evolution of the radius of a 0.36\,M$_{\rm J}$ planet made
  of a central core of 30 to 75\,M$_\oplus$ and a gaseous envelope of
  solar composition. The planet is irradiated so that $T_{\rm
  eq}^\ast = 1740\,$K (see text). The core is assumed to be made of ices
  (plain lines) or rocks (dashed lines). The results are compared to
  the observational constraints on the age and radius of HD149026b (black box).}
\label{fig:evol_std}
\end{figure}

The calculations are made for various core masses and core
compositions and compared to the observational constraints in
Fig.~\ref{fig:evol_std}. As discussed by \citet{Sato05} and by
\citet{Fortney06}, the small planetary radius implies that it
contains a large fraction of its mass in heavy elements. 
Difference in composition of the core---whether these heavy elements are mostly
``ices'' (a mixture of water, ammonia and methane in their [unknown]
high pressure, relatively high temperature form) or ``rocks'' (a
mixture of refractory materials including mostly silicates)---affects the
radius of the planet by 10 to 20\%. For comparison, the
interiors of Uranus and Neptune are consistent with being mostly made
of ``ices'' \citep[e.g.,][]{PPM00}. 
Figure~\ref{fig:evol_std} thus confirms the need for a substantial amount of
heavy elements in HD149026b when adopting a standard evolution
scenario. 
Quantitatively, this implies core masses between 41 and
83\,M$_\oplus$ in good agreement with \citet{Sato05} and
\cite{Fortney06}. In particular, the latter yields values between 60
and 93\,M$_\oplus$. The difference is due to the fact that in order to
constrain more strictly the minimum mass of heavy elements present in
the planet, we used a low value of the atmospheric temperature.

\subsection{The Cold Storage Hypothesis \label{sec:cold_storage_hypothesis}}

We now investigate the hypothesis that the planet may have cooled
far from its parent star before suddenly being sent into the presently
observed 0.046~AU orbit. The reason for this sudden event is to be
determined, but could be due to dynamical planet-planet interactions
(see section \ref{sec:late_Z_supply}). 
In terms of evolution, the planet is allowed to cool more
rapidly during the time-period when it is far from the parent
star. The sudden inward migration indeed yield an expansion of the
planet but it may be limited to the outer layers and thus be
relatively small \citep{Burrows00}. We want to investigate whether
the present mass and radius are compatible with a relatively small
core. 

We thus calculate evolution models similar to those in the
previous section, but using $T_{\rm eq}=100\,$K and an atmospheric
model based on simplified radiative transfer calculations of an
isolated atmosphere \citep[see][]{Saumon96,Guillot05}. Based on
these calculations, we obtain generally a much faster contraction of the planet; 
a less than 30\,$\mearth$ core can account for the observed radius. 
However, we have to include 
reheating of the outer shell of the envelope. This shell is
approximatively defined by the region whose temperature is lower than
the new equilibrium temperature, 1740\,K. Based on our calculations,
this corresponds to a pressure level of $\sim$10\,kbar.

A simple estimation of the reheating timescale of the outer shell 
suggests that we have to include this effect. 
Using equations of radiative diffusion and energy conservation, 
we estimate that the reheating timescale of a layer of pressure $P$,
temperature $T$, opacity $\kappa$ and heat capacity $c_p$ is to first
order independent of its initial temperature and equal to:
\begin{equation}
\tau_{\rm rad}\approx {3c_p\over 4ac}{\kappa P^2\over g^2 T^3},
\end{equation}
where $a$ is the radiation density constant, $c$ the speed of light, and $g$
the planet's gravity (assumed uniform). This expression is consistent
with direct simulations by \citet{IBG05}.
This timescale is about $10^8\,$years for
$\kappa\approx 1\,\rm cm^2\,g^{-1}$, $P\approx 10$\,kbar, 
  $g\approx 1550\rm\,cm\,s^{-2}$, $T\approx 2000\,{\rm K}$ and $c_p\approx 8\times
  10^7\rm\,erg\,g^{-1}\,K^{-1}$,
which implies that the reheating is relatively fast and should involve most
of the shell that we defined. 

However, the expansion due to the reheating is relatively small. 
The $P^2$ dependence ensures that most of this outer shell will be affected, 
and it will expand by a radius estimated to be:
\begin{equation}
\Delta R\approx H_p\Delta\ln P, 
\end{equation}
where $H_p\approx{\cal R}T/\mu g$ is the new pressure scale height and
 $\Delta\ln P$ measures the depth of the expanding shells in pressure
units. Based on our simulations, a rough estimate is that the planet
should expand by $\sim 4\times 10^8\,$cm ($= 0.06 R_{\rm J}$) 
when considering levels between 10\,kbar and 1 bar. 

Note that this estimate does not include factors such as global
reheating due to tidal dissipation caused by the necessary circularization
of the planet's spin and eccentricity. 
If such dissipation occurs
sufficiently deep in the planet, this will lead to an expansion of the
planet's deeper layers that could completely erase the planet's cold
evolution phase \citep[see][and references therein]{Guillot05}. Thus, this
very possibility that a planet can contract more efficiently far away
from the star and then be brought in with little consequence on its
radius is very much in question. We choose however to consider it
because it gives a sense of the robustness of the conclusion that
HD149026b contains a large fraction of heavy elements.
\begin{table}[htbp] 
\caption{Derived constraints on HD149026b's composition. 
$M_{\rm H\,He}$ and $M_Z$ are masses of H/He and  
heavy elements, and $M_{\rm p} = M_{\rm H\,He} + M_Z$ is 
total planetary mass. 
} 
\label{tab:constraints} 
\begin{center} 
\begin{tabular}{cccc} 
\hline\hline 
& $M_Z \rm\ (M_\oplus)$ & $M_{\rm H\,He}\rm\ (M_\oplus)$ &  
$M_Z/M_{\rm p}$\\ \hline 
\multicolumn{4}{l}{{\it In situ evolution}} \\ 
\ \ ices  & $50-83$ & $27-54$ & 0.52--0.72 \\ 
\ \ rocks & $41-69$ & $38-64$ & 0.43--0.60 \\\hline 
\multicolumn{4}{l}{{\it Cold storage scenario}} \\ 
\ \ ices  & $37-77$ & $34-66$ & 0.39--0.67 \\ 
\ \ rocks & $33-61$ & $46-76$ & 0.33--0.53 \\\hline 
\multicolumn{4}{l}{{\it Total}} \\ 
          & $33-83$ & $27-76$ & 0.33--0.72 \\\hline\hline 
\end{tabular} 
\end{center} 
\end{table} 

Adding $\Delta R$ ($\sim 4 \times 10^8$ cm) 
to the radius obtained by the evolution calculation of the isolated planet 
(not including the global reheating), 
we derive masses of heavy elements that
reproduce the observed radius and show these in
Table~\ref{tab:constraints}. Even in the case of a long storage 
($\ga 10^8$ years) of the
planet in a cold environment (corresponding to more than 10~AU), a
large amount of heavy elements (strictly more than 33\,M$_\oplus$)  is
required to reproduce the observed radius of HD149026b.

\subsection{Heavy Elements: In the Core or in the Envelope? 
\label{sec:core_or_envelope}}

We have thus far assumed all heavy elements to lie within a central
core. In order to see how the results are dependent on this
assumption, we compare our previous model with a 60\,M$_\oplus$ ice core
to a similar model with a 30\,M$_\oplus$ ice core and an envelope which
is enriched by 30\,M$_\oplus$ of ices. For the ices in the envelope,
we use the EOS described by \citet{SG04}. 

\begin{figure}[htbp]
\plotone{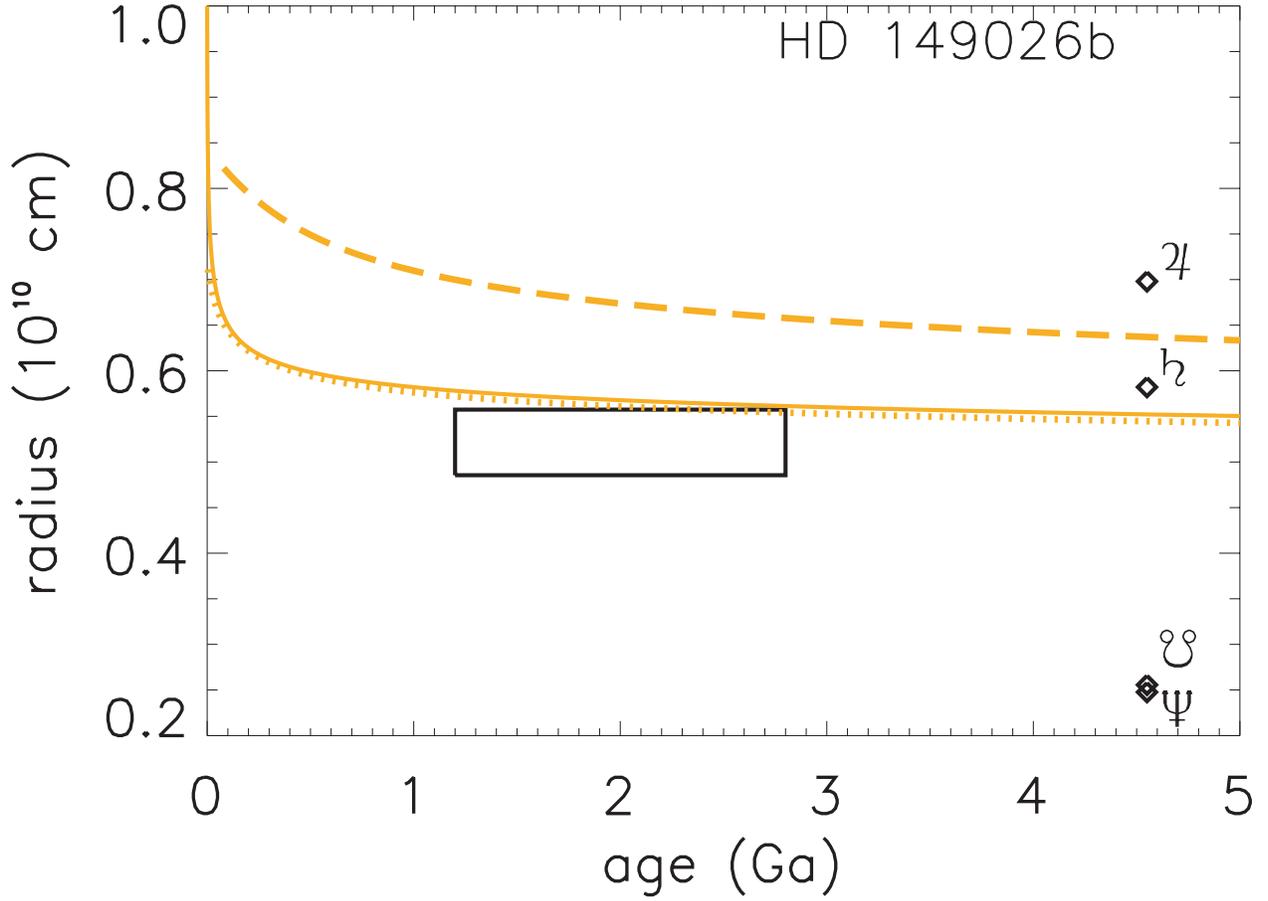}
\caption{Evolution of the radius of a 0.36\,M$_{\rm J}$ planet. The
  plain line corresponds to the case of a 60\,M$_\oplus$ ice core with
  a solar composition envelope shown in Fig.~\ref{fig:evol_std}. The
  dotted line represents the result of the calculation with a 30\,M$_\oplus$ ice core
  and 30\,M$_\oplus$ of ices in the envelope; opacities were calculated
  for a solar-composition mixture. The dashed line is the same
  calculation, but with an opacity that is artificially increased by
  30 to account for the presence of heavy elements in the envelope.
  The results are compared to the observational 
  constraints on the age and radius of HD149026b (black box).}
\label{fig:evol_opa}
\end{figure}

Figure~\ref{fig:evol_opa} shows the results of the calculations in the two cases: 
(i) with unchanged opacities (i.e., as calculated for a
  solar-composition mixture) and 
(ii) with 30 times larger opacities to mimic the effect of the
  enrichment of heavy elements in the envelope.
Basically, when using an unchanged opacity table (the dotted line), 
there is very little difference 
between a planet that has all its heavy elements in the core 
and a planet that has them mixed throughout its envelope. 
However, large differences arise when the opacities are
affected proportionally to the amount of heavy elements that are mixed 
(the dashed line). 
In that case, the cooling and contraction timescale, which is
dominated by radiative transport in the outer radiative zone, becomes
long, and prevents the rapid contraction of the planet. 

These results confirm again that 
our estimates of the amount of heavy elements in HD149026b are lower
limits and that the planet indeed must contain a significant amount 
of heavy elements.
If most of the heavy elements are located below the radiative zone, 
which is not unlikely,
then we expect our constraints to be relatively accurate. On the other
hand, if the material is mixed in the outer radiative zone, 
we expect that its interior contains more heavy elements by up to
$\sim 20\,\rm M_\oplus$ than calculated here.

\subsection{Impact of the Various Parameters \label{sec:various_parameters}}

We have used a rather simplified approach for the calculations 
of the interior structure. 
On the other hand, reality is without doubt more complex. 
However, this should not affect significantly the global constraints 
that are derived in Table~\ref{tab:constraints}. 
First, the two extreme compositions used for the heavy elements
(``ices'' and ``rocks'') ensure 
that most variations due to improved EOS, for example, 
are likely to fall in between the range of values that are considered. 

Our external boundary condition is extremely simplified. Although it
agrees with more detailed atmospheric models, one has to account for
the fact that the atmospheric temperatures may be higher, in
particular if the atmosphere is itself enriched in heavy elements 
\citep[see][]{Fortney06}. A higher atmospheric temperature leads to a
slightly larger radius, everything else being the same. However, given
the decrease of the opacity with increasing temperature in that
regime, the radius increase is very limited. 

Other factors point toward a slightly larger amount of heavy elements
than calculated here: (i) as discussed, the likely increase of the
opacities in an enriched envelope; (ii) the presence of any other
energy source such as the one that is required to reproduce the radius
of HD209458b; (iii) thermally-dependent EOS for the
core. Again, given that the core mass is already quite large, we
expect these factors not to change the masses of heavy elements that
are inferred above.

\section{A SUBCRITICAL CORE ACCRETION SCENARIO? \label{sec:accretion}}

The structure of HD149026b as derived in the previous section is
certainly puzzling: It contains as much hydrogen and helium as
Saturn, but at least twice as much heavy elements. In other words, the
ratio of heavy elements to hydrogen and helium is intermediate between
those of Saturn and Uranus/Neptune. 
In the core accretion scenario, 
when core mass exceeds a critical value (called the \textit{critical core mass}), 
the Kelvin-Helmholtz contraction of the envelope takes place, 
resulting in substantial disk-gas accretion onto the planet 
\citep[e.g.,][]{Mizuno80,BP86}. 
A widespread idea is that Jupiter and Saturn experienced 
the substantial gas accretion beyond the critical core mass, 
while the cores of Uranus and Neptune remained subcritical 
because of progressive accretion of planetesimals 
until the gaseous circumstellar disk disappeared \citep{P96}. 
In this section, we investigate
whether HD149026b may have formed in a way similar to what envisioned
for Uranus and Neptune.

\subsection{Critical Core Mass and Gas Accretion Rate \label{sec:critical_mass}}

\begin{figure}
\plotone{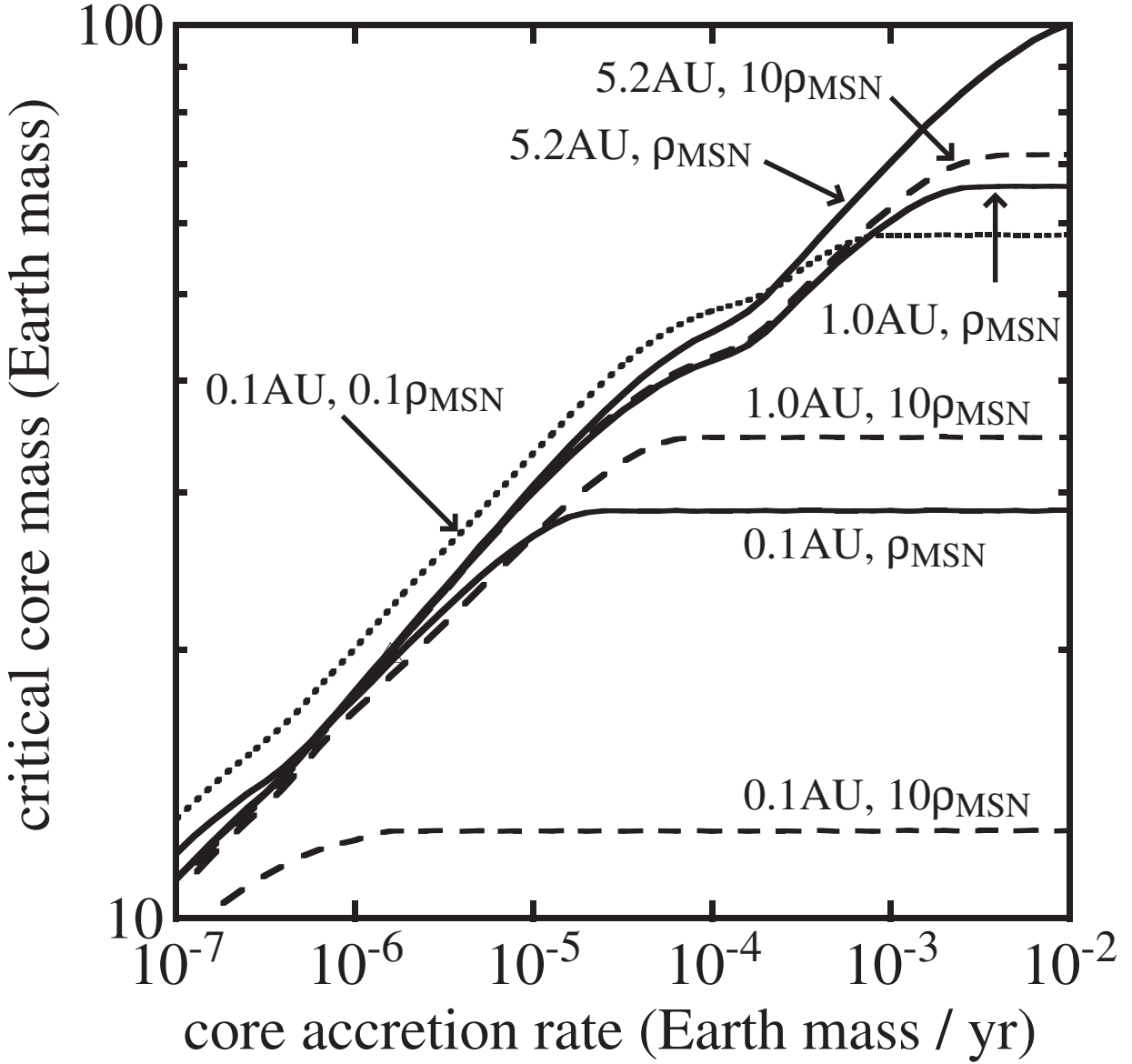}
\caption{
The critical core mass as a function of core accretion rate for 
different choices 
of distance from the parent star and local density of disk gas. 
}
\label{fig:critical_core_mass}
\end{figure}

Figure \ref{fig:critical_core_mass} shows the critical core mass 
($M_{\rm c,crit}$) as a function of core accretion rate 
for several different choices of distance from the parent star 
and local density of disk gas. 
In the numerical simulations 
we have used the H-He EOS from \citet{SCVH95}, 
the grain opacity from \citet*{PMC85}, and 
the gas opacity from \citet{AF94}. 
The integration method is basically the same as that of \citet{Ikoma00}
who used simpler forms of EOS and gas and grain opacities. 

The input physics used here is somewhat different from 
that used in section~\ref{sec:present_structure}. 
First we include grain opacity that was not included in the simulations 
in section 2, because an accreting envelope contains low temperature parts 
in which grain opacity is dominant, 
unlike the fully-formed planet ($T \ge 1700$~K). 
Although the fully-formed planet also contains such low temperature parts 
in the cold storage model, the planet is almost fully convective. 
Second the gas opacity used in this simulation is different from 
that used in the previous section. 
This difference has little influence on the results obtained in this section, 
because most of the part of the envelope where gas opacity is dominant relative to 
grain opacity is convective. 
Finally, although constant core density is assumed here, 
the critical core mass is known to be insensitive to core density \citep{MNH78}.

Figure~\ref{fig:critical_core_mass} shows 
a large subcritical core can be formed 
if core accretion rate is high. 
High core accretion rates stabilize the envelope, 
yielding progressively larger critical core masses \citep{Ikoma00}. 
This, however, remains true 
up to a point for which the envelope becomes fully convective; 
the structure of a fully convective envelope is almost adiabatic 
and thus independent of energy flux supplied by incoming planetesimals 
(i.e., core accretion rate). 
At that point, the critical core mass reaches its maximum value. 
In this case, this {\it maximal} critical core mass depends 
on the local disk conditions 
(i.e., the local density and temperature of disk gas and the distance from the
parent star) \citep*{Wuchterl93,Ikoma01}. 
In the minimum-mass solar nebula (MSN) proposed by \citet{Hayashi81}, 
the entropy of the disk gas decreases as the distance to the parent star decreases. 
That is why the critical core mass is smaller at locations closer to the
parent star: 
In a denser gas disk, the critical core mass is smaller for the same reason 
\citep{Ikoma01}.
As a result, formation of a large subcritical core of 50--80~$\mearth$ 
requires not only high core accretion rate, 
but also a location not too close to its parent star 
and not too massive gaseous disk. 

\begin{figure}
\plotone{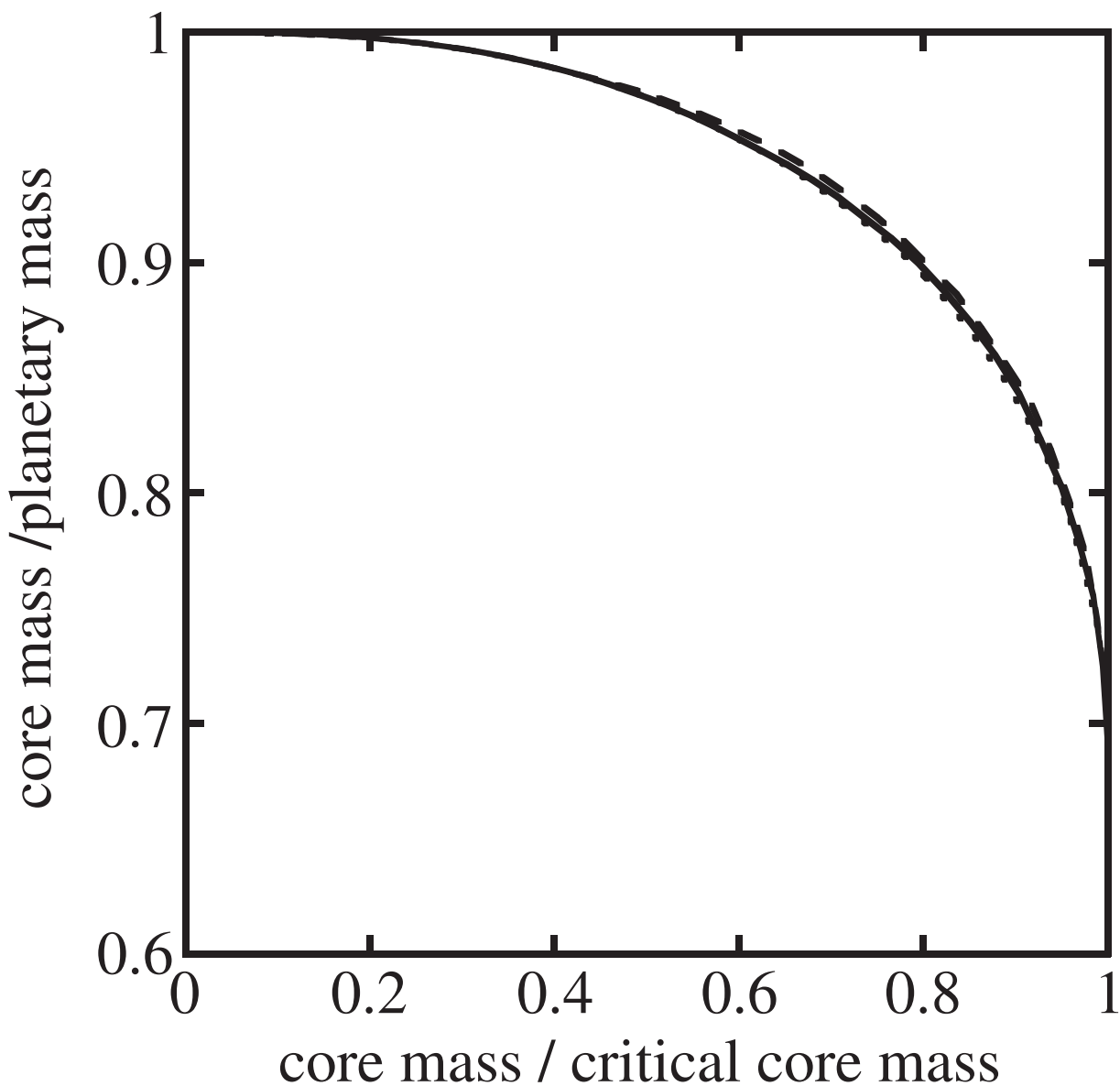}
\caption{
The ratio of core mass to planetary total mass 
as a function of core mass normalized by the critical core mass. 
The dotted line represents the result for $\dot{M}_{\rm c}$ of 
$1 \times 10^{-5} M_\oplus / {\rm yr}$ at 0.1~AU, 
the dashed one for $\dot{M}_{\rm c}$ of 
$1 \times 10^{-5} M_\oplus / {\rm yr}$ 
at 1~AU, and the solid one for $\dot{M}_{\rm c}$ of 
$1 \times 10^{-3} M_\oplus / {\rm yr}$ at 5.2~AU. 
}
\label{fig:critical_ratio}
\end{figure}

Although a large core of 50--80 $\mearth$ can, in principle, be formed 
by subcritical growth as shown above, 
the ratio of the critical core mass to the corresponding
total (core + envelope) mass suggests that the core mass of 
HD149026b was likely to be supercritical.  Figure
\ref{fig:critical_ratio} shows the ratio of core mass to planetary
total mass as a function of core mass (normalized by the critical core
mass) for three different sets of core accretion rate and distance
from the parent star.  This ratio is found to be rather insensitive to 
values of the parameters; it is always $\sim 0.7$ at
the critical core mass.  The evolution model of HD149026b obtained in
section \ref{sec:present_structure} shows that the current ratio of
core mass to planetary mass is smaller in most cases (see
Table~\ref{tab:constraints}).  Although there are a few cases where
the current ratio is as large as $\sim 0.7$, the probability of the
formation of such planets seems to be low.  
From Fig.~\ref{fig:critical_ratio} we also find the duration during which
the ratio of core mass to planetary mass is around $\sim 0.7$ is quite
short in the total core formation time.
It follows from this consideration that HD149026b is likely to have experienced 
disk-gas accretion due to the Kelvin-Helmholtz contraction of the envelope.

\begin{figure}
\plotone{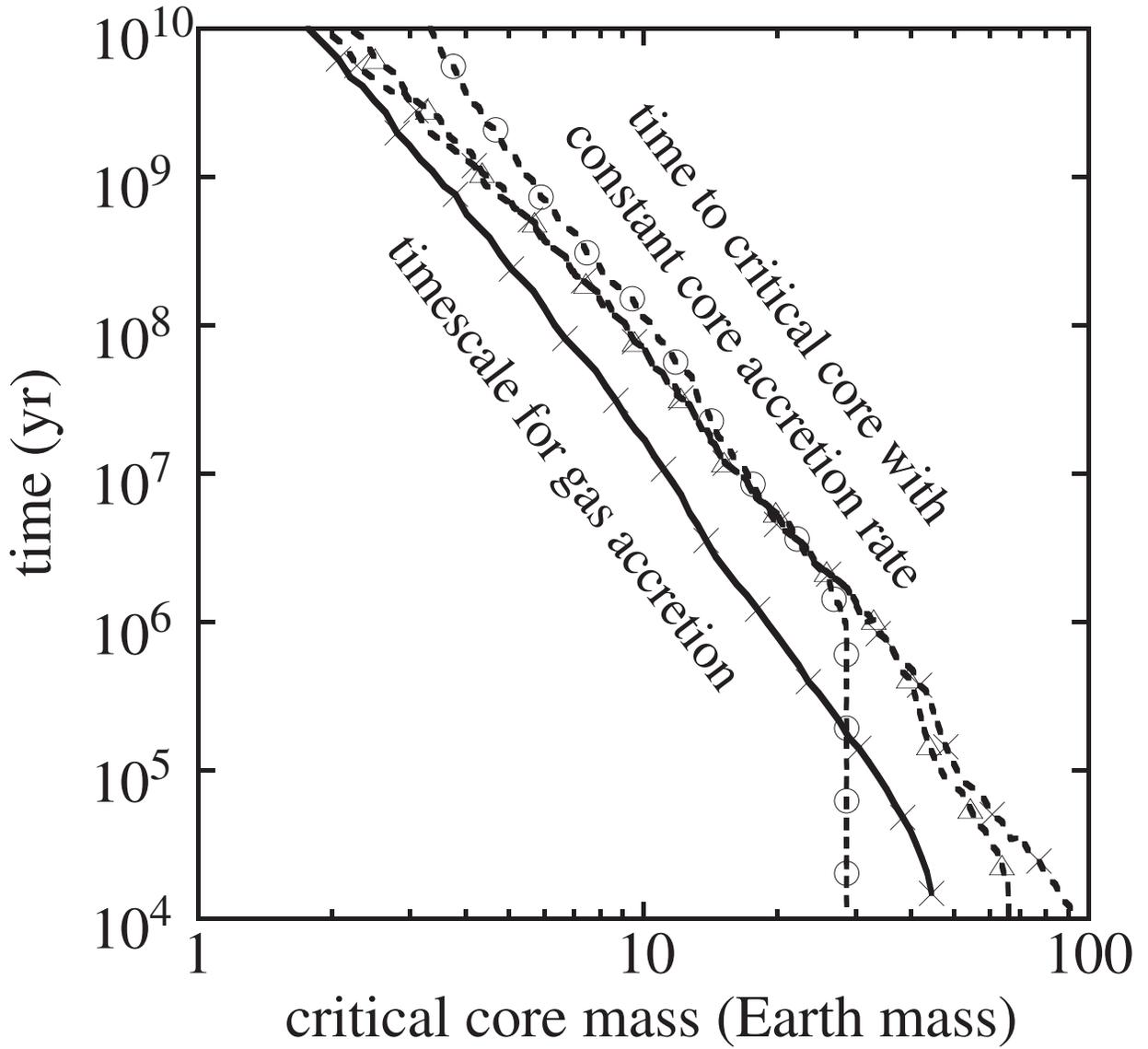}
\caption{
Timescale for gas accretion (solid line) 
is shown as a function of the critical core mass 
together with time to the critical core (dashed lines). 
The timescale for gas accretion is that for an $e$-fold increase 
in envelope mass just after the critical core mass is attained 
(see text for the exact definition). 
The circles, triangles, and crosses are for 
0.1, 1, and 5.2~AU in the minimum-mass solar nebula.
The vertical part of the dotted curve with circles corresponds to 
the horizontal part of the corresponding line 
in Fig.~\ref{fig:critical_core_mass}.
}
\label{fig:KH}
\end{figure}

The rate of the gas accretion 
determined by the Kelvin-Helmholtz contraction of the envelope 
beyond the critical core mass---the gas accretion is hereafter called 
the KH gas accretion---depends 
strongly on the critical core mass. 
Figure~\ref{fig:KH} shows the typical timescale ($\tau_{\rm g}$) 
for the KH gas accretion defined by 
\begin{equation}
\tau_{\rm g} = C \frac{G M_{\rm c,crit} M_{\rm e,crit}}
                           {R_{\rm conv} L},
\label{eq: KH time}
\end{equation}
where $G$ is the gravitational constant, 
$M_{\rm c,crit}$ is the critical core mass, 
$M_{\rm e,crit}$ is the corresponding envelope mass, 
$R_{\rm conv}$ is the outer radius of the inner convective region, and 
$L$ is the luminosity \citep{Ikoma00}. 
The numerical factor, $C$, is chosen to be $3/2$ 
based on the numerical simulations by \citet{IG06}. 
In cases where core accretion is suddenly stopped 
(corresponding to core isolation; see below), 
$C$ should be $1/3$ \citep{IG06}. 
On this timescale the envelope mass increases by a factor of $e$. 
The timescale for the KH gas accretion is approximately expressed by 
\begin{equation}
\tau_{\rm g} \sim 5 \times 10^{10} 
\left(\frac{M_{\rm c,crit}}{\mearth} \right)^{-3.5}
{\rm years}
\label{eq:KHtime_formula}
\end{equation}
in this range\footnote{
The form of equation~(\ref{eq:KHtime_formula}) is different from 
that of equation~(19) of \citet{Ikoma00}. 
The difference comes mainly from the difference in grain opacity 
\citep[see][for the detailed discussion]{IG06}.
}. 
This relation implies that if the core mass is small, 
the timescale of the KH gas accretion is much longer than the disk lifetime 
($\sim 10^6$-$10^7$ years). 

For discussion in section \ref{sec:subcritical}, 
we have also drawn lines representing 
the time for formation of a critical core with constant core accretion rate, 
\begin{equation}
\tau_{\rm c,acc,crit} = \frac{M_{\rm c}}{\dot{M}_{\rm c}} \mid_{M_{\rm c}=M_{\rm c,crit}},
\label{eq:core_growth_time}
\end{equation}
for three different sets of the parameters.

\subsection{Subcritical Core Accretion \label{sec:subcritical}}

We examine the subcritical core accretion 
to know how large core can be formed 
before the onset of the KH gas accretion. 
Core mass is limited by the isolation mass 
that is determined by the total mass of solid components in the core's feeding zone,
unless planetesimals are supplied from outside the feeding zone
by migration of planetesimals or the core itself.
Merging between the isolated cores is not likely until
disk gas is severely depleted 
\citep{Iwasaki02, Kominami02}, because the eccentricity damping due 
to disk-planet interaction \citep{Artymowicz93, Ward93} is
too strong to allow orbital crossing of the isolated cores.
The core isolation mass is thus given by \citep{IL04a}
\begin{equation}
M_{\rm c,iso} \simeq 0.16 f_{\rm d}^{3/2} \eta_{\rm ice}^{3/2}
                    \left(\frac{a}{1{\rm AU}}\right)^{3/4} 
                    \left(\frac{M_*}{{\rm M}_{\odot}}\right)^{-1/2} \mearth,
\label{eq:M_c_iso}
\end{equation}
where $a$ is semimajor axis of the core.
The scaling factor $f_{\rm g}$ and
$f_{\rm d}$ express enhancement of 
disk surface density of gas ($\Sigma$) and dust ($\sigma$)
components relative to those of MSN, defined by
\begin{equation}
\begin{array}{lll}
\Sigma & = & 
\displaystyle{2400 f_{\rm g} \left(\frac{a}{1{\rm AU}}\right)^{-3/2} {\rm g/cm}^2} \\
\sigma & = & 
\displaystyle{10 f_{\rm d} \eta_{\rm ice} 
              \left(\frac{a}{1{\rm AU}}\right)^{-3/2} {\rm g/cm}^2}. 
\end{array}
\end{equation}
In the solar metallicity cases,
the values of these factors may have a range between 0.1 and 10 \citep{IL04a}.
For metal-rich disks, the range of $f_{\rm d}$ may shift to larger values by
a factor of $10^{\rm [Fe/H]}$, while that of $f_{\rm g}$ does not change
\citep{IL04b}.
$\eta_{\rm ice}$ expresses an enhancement of disk surface
density due to ice condensation, which is 3--4 beyond an ice line
at $\sim 3$~AU. 
Equation~(\ref{eq:M_c_iso}) shows large cores form at large $a$,
in particular beyond the ice line, in disks with large $f_{\rm d}$.
As seen from Fig.~\ref{fig:critical_core_mass}, 
smaller core accretion rate yields smaller critical core mass, 
which implies core isolation triggers the KH gas accretion. 
For a core of $M_{\rm c}$ not to be isolated, that is, 
$M_{\rm c} < M_{\rm c, iso}$, 
\begin{equation}
f_{\rm d} > 30 \left(\frac{M_{\rm c}}{30 \mearth}\right)^{2/3} 
            \eta_{\rm ice}^{-1} \left(\frac{a}{1{\rm AU}}\right)^{-1/2} 
            \left(\frac{M_*}{{\rm M}_{\odot}}\right)^{1/3}.
\label{eq:f_iso}
\end{equation}

Next, we consider the condition for a core to avoid the KH gas accretion 
until its isolation.  
Figure 3 shows that $M_{\rm c,crit}$ is expressed approximately by 
\begin{equation}
M_{\rm c,crit} 
\simeq 10 \left(\frac{\dot{M}_{\rm c}}
                     {10^{-7}\mearth/{\rm yr}}
          \right)^{0.2} \mearth,
\label{eq:m_crit}
\end{equation}
except for fully convective cases. 
Note that consideration of fully convective cases does not change 
our conclusion (see below). 
The condition for a core to avoid the KH gas accretion is 
$M_{\rm c} < M_{\rm c, crit}$. 
Using $\tau_{\rm c,acc} \equiv M_{\rm c}/\dot{M}_{\rm c}$ and
Eq.~(\ref{eq:m_crit}), 
this condition is rewritten by (see Fig.~\ref{fig:KH})
\begin{equation}
\tau_{\rm c,acc} < 
\tau_{\rm c,acc,crit} \simeq 1 \times 10^{6}
\left(\frac{M_{\rm c}}{30 \mearth}\right)^{-4}{\rm years}.
\label{eq:m_crit_time}
\end{equation}
For cores of 30, 45 and $70 {\rm M}_{\oplus}$,
$\tau_{\rm c,acc}$ must be shorter than $1 \times 10^{6}$ years, 
$2 \times 10^5$ years, and $3 \times 10^4$ years, respectively. 
The core growth timescale due to planetesimal accretion 
is given by \citep{IL04a}
\begin{equation}
\tau_{\rm c,acc} \simeq 9.3 \times 10^5 f_{\rm d}^{-1} f_{\rm g}^{-2/5} 
                    \eta_{\rm ice}^{-1}
                    \left(\frac{a}{1{\rm AU}}\right)^{27/10} 
                    \left(\frac{M_{\rm c}}{30 \mearth}\right)^{1/3} 
                    \left(\frac{M_*}{{\rm M}_{\odot}}\right)^{-1/6} 
                    \left(\frac{m}{10^{21} \rm g}\right)^{2/15}
                    \mbox{years},
\label{eq:tau_c_acc}
\end{equation}
where $m$ is the mass of a planetesimal. 
The dependence on $f_{\rm g}$ appears because 
damping of velocity dispersion of planetesimals 
due to aerodynamical drag is taken into account.
Substituting Eq.~(\ref{eq:tau_c_acc}) into Eq.~(\ref{eq:m_crit_time}), 
we have 
\begin{equation}
f_{\rm d} > 0.95 \left(\frac{M_{\rm c}}{30 \mearth}\right)^{65/21} 
            \left(\frac{f_{\rm d}}{f_{\rm g}}\right)^{2/7} 
            \eta_{\rm ice}^{-5/7} \left(\frac{a}{1{\rm AU}}\right)^{27/14} 
            \left(\frac{M_*}{{\rm M}_{\odot}}\right)^{-5/42}
            \left(\frac{m}{10^{21} \rm g}\right)^{2/21}.
\label{eq:f_tau}
\end{equation}
This inequality indicates that large $f_{\rm d}$ is needed 
to keep core accretion rate high enough 
to avoid the KH gas accretion. 

\begin{figure}[htbp]
\plotone{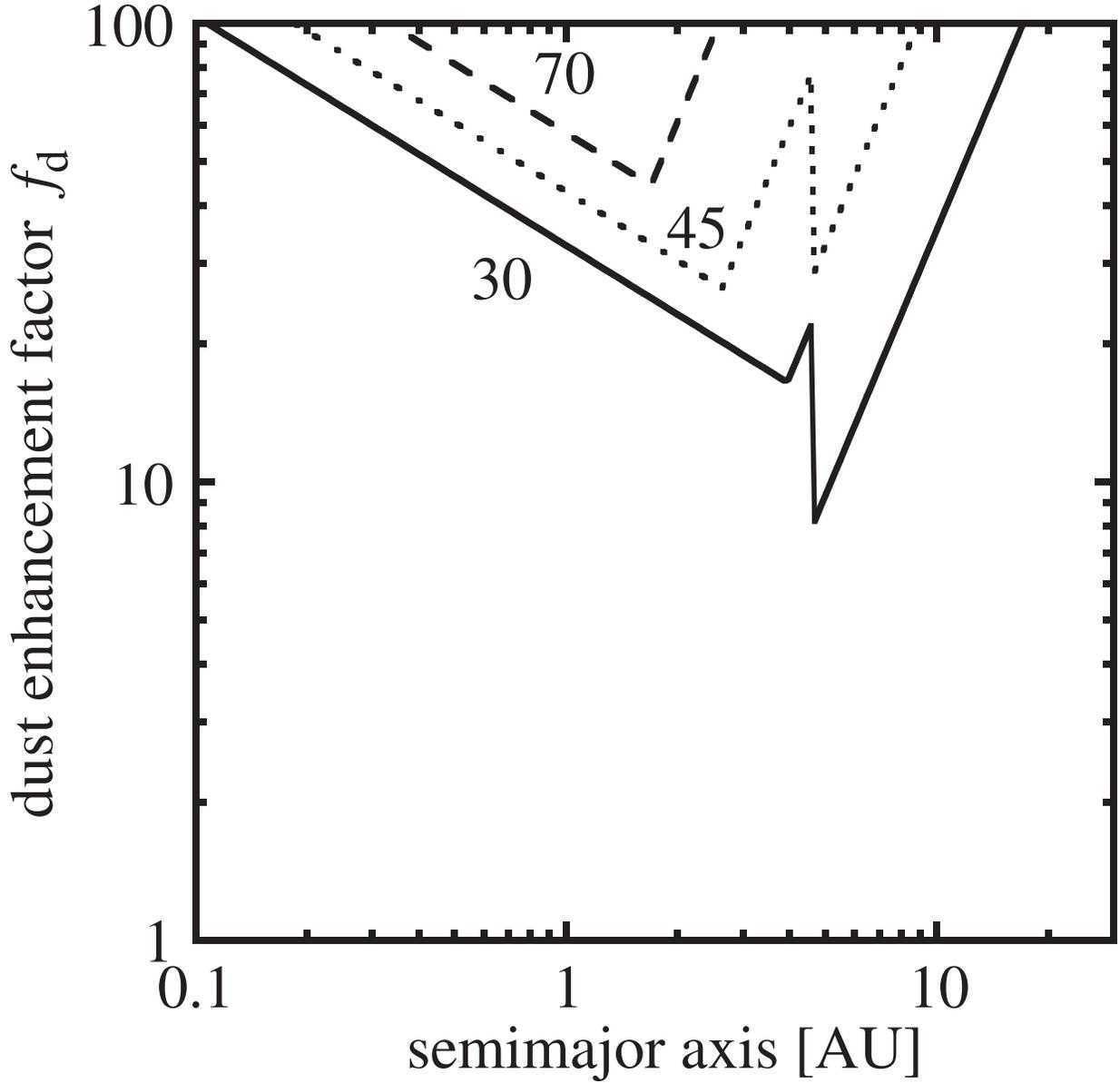}
\caption{
The conditions for subcritical core accretion.
The subcritical core accretion is possible in the
regions over the solid, dotted, 
and dashed lines for $30, 45$ and $70{\rm M}_{\oplus}$ cores,
respectively. 
Corresponding to HD149026, $M_* = 1.3M_\odot$ and [Fe/H]=0.36 
($f_{\rm d}/f_{\rm g} = 2.3$) are assumed.
Note that the discontinuities of the curves come from 
sudden increases in disk surface density of dust component 
at the snow line.
}
\label{fig:condition}
\end{figure}

For subcritical core accretion,
both Eqs.~(\ref{eq:f_iso}) and (\ref{eq:f_tau}) must be satisfied.
For $M_* = 1.3 {\rm M}_\odot$ and [Fe/H]=0.36 ($f_{\rm d}/f_{\rm g} = 2.3$), 
which correspond to HD149026, the conditions are plotted
in Fig.~\ref{fig:condition}.
Subcritical core accretion could be possible far from the parent star 
in relatively massive disks for $30{\rm M}_{\oplus}$ cores.
However, for $45$ and $70{\rm M}_{\oplus}$ cores,
extremely massive or extremely metal-rich disks with 
$f_{\rm d} > 30-50$ are required. 
Even for those large values of $f\sub{d}$, 
since the critical core mass reaches its maximal value 
in cases of high core accretion rate, 
subcritical formation of 45 and 70~$\mearth$ is impossible in some cases 
(see Fig.~\ref{fig:critical_core_mass}).
Although formation of a $30{\rm M}_{\oplus}$ core may 
not be very difficult in relatively metal-rich disks, 
$30{\rm M}_{\oplus}$ is the lowest estimated value 
for $M_Z$ of HD149026b in the cold storage model; 
$50$--$80{\rm M}_{\oplus}$ is most likely.
Therefore, heavy elements must be supplied
after substantial accretion of the envelope.  
This issue is discussed in section \ref{sec:Z_supply}.

\section{SEVERAL POSSIBILITIES FOR THE FORMATION OF HD149026b 
\label{sec:possibilities}}

The investigations in sections~\ref{sec:present_structure} and \ref{sec:accretion}
have revealed that we have to explain at least three properties of HD149026b:
its small orbital radius, its high metalicity, 
and its moderate amount of hydrogen and helium. 

A possible scenario that explains the proximity of the planet to its parent star 
is the {\it in situ} formation or migration of the planet. 
The {\it in situ} formation is however unlikely, 
because the core isolation mass is so small at $\sim$ 0.04~AU 
(see eq.~[\ref{eq:M_c_iso}]) 
that the planet can get only a tiny amount of hydrogen/helium 
within the disk lifetime 
because of too long timescale for the KH gas accretion 
(see eq.~[\ref{eq:KHtime_formula}]).

Section~\ref{sec:subcritical} has shown cores of $\ga 30 \mearth$ can form 
only far from the parent star in reasonably massive and/or metal-rich disks. 
Except for the cold storage model (section \ref{sec:cold_storage_hypothesis}),
residual heavy elements of 20--$50{\rm M}_{\oplus}$ must be
supplied to the planet. 
In section~\ref{sec:Z_supply}, 
we examine two possible scenarios; 
supply of heavy elements (planetesimals or another planet) 
during gas accretion (\S~\ref{sec:concurrent_Z_supply}) and 
after disk gas depletion (\S~\ref{sec:late_Z_supply}). 
To account for the moderate amount of hydrogen and helium 
of HD149026b, 
we examine limited supply of disk gas (\S~\ref{sec:limited_gas_supply}) and 
loss of the envelope gas (\S~\ref{sec:envelope_loss}).

\subsection{Supply of Heavy Elements \label{sec:Z_supply}}

\subsubsection{During gas accretion \label{sec:concurrent_Z_supply}}
First we briefly discuss the possibility 
that the large amount of heavy elements in HD149026b were supplied 
\textit{during} substantial accretion of the envelope 
that happened far from the parent star. 
An increase in the planet's mass due to gas accretion expands its feeding zone. 
If the planet can efficiently capture the planetesimals in the feeding zone, 
supply of heavy elements during gas accretion is possible \citep*{P96,Alibert04}. 
However, the cross section of gravitational scattering is much larger 
than that of collision in outer regions. 
Most planetesimals are thus scattered before accretion onto the planet. 
The gravitational scattering combined with eccentricity damping 
by disk-gas drag ends up opening a gap in the planetesimal disk: 
This phenomenon is called ``shepherding'' \citep[e.g.,][]{TI97}\footnote{
The shepherding by isolated mean-motion resonances is called 
``resonant trapping'' \citep[e.g.,][]{WD85, KLG93}. }.
Although fast migration or growth of the planet can avoid the gap opening, 
unrealistically fast migration or growth may be required 
for a Saturn-mass planet to avoid the gap opening \citep{TI99, Ida00}. 
Thus, supply of a large amount of planetesimals may be unlikely 
to happen during gas accretion. 

However, much more detailed studies are needed for this possibility. 
If the shepherding completely prevented the planetesimal accretion 
throughout the gas accretion phase, 
the well-known enrichment of heavy elements in the envelopes of Jupiter and Saturn 
would be unable to be accounted for. 
A possibility is accretion of small fragments. 
Small fragments are produced by disruptive collisions of planetesimals 
in the vicinity of a large core \citep{IWI03}.
Since atmospheric-gas drag is strong for the fragments, 
they are accreted onto the core \citep{II03}. 
Because these studies are limited to the subcritical core case, 
we need to extend the calculation to the case of supercritical cores, 
in which the planet mass rapidly increases and 
non-uniformity of disk gas is pronounced. 

\subsubsection{After disk gas depletion \label{sec:late_Z_supply}}
We next explore the possibility of 
enrichment of heavy elements in HD149026b 
(i.e., bombardment of another planet or planetesimals) 
\textit{after} substantial accretion of the envelope. 

If the late bombardment happened, it is likely to have occurred 
after the planet migrated to its current location: 
The close-in planet collided with another planet(s) or planetesimals 
that had been gravitationally scattered by an outer giant planet(s). 
At $\sim$ 0.05~AU 
the planet's Hill radius is only several times as large as 
its physical radius, while the former is much larger 
(by 100--1000) than the latter in outer regions ($\ga 1$~AU). 
That means the ratio of collision to scattering cross sections is much larger 
at $\sim$ 0.05~AU relative to in outer regions.  
The scattering cross section is further reduced by high speed encounter. 
The impact velocities ($v_{\rm imp}$) of the scattered bodies 
in nearly parabolic orbits are as large as the local Keplerian velocity 
at $\sim$ 0.05~AU, which is a few times larger than
surface escape velocity of the inner Saturn-mass planet ($v_{\rm esc}$).
The collision cross section is, in general, larger than 
the scattering cross section for $v_{\rm imp} > v_{\rm esc}$.
Note that the shepherding does not work for
highly eccentric orbits.

To know how efficiently such scattered bodies collide with 
the close-in giant planet, 
we perform the following numerical simulation. 
We consider an inner planet of $0.5M_{\rm J}$ 
in a circular orbit at 0.05~AU 
and a planetesimal (test particle) or another giant planet of $0.5M_{\rm J}$ 
in a nearly parabolic orbit of $e \simeq 1$ and $i = 0.01$ 
with initial semimajor axis of $a = 1$~AU. 
For each initial pericenter distance ($q$), 100 cases with 
random angular distributions are numerically integrated 
by 4th order Hermite integrator for 100 Keplerian periods at 1~AU. 
Then, collision probability with the inner planet ($P_{\rm col}$), 
that with the parent star ($P_{\rm col}^*$), 
and ejection probability ($P_{\rm ejc}$) are counted. 
The residual fraction, $1 - (P_{\rm col} + P_{\rm col}^* + P_{\rm ejc})$, 
corresponds to the cases in which the planetesimal/planet 
is still orbiting after 100 Keplerian periods. 
We also did longer calculations 
and found that $P_{\rm col}/P_{\rm ejc}$ is similar, although
individual absolute values of $P_{\rm col}$ and $P_{\rm ejc}$ increases.
The results are insensitive to $a$ as long as $a \gg 0.05$~AU, 
because velocity and specific angular momentum of the incoming 
planetesimal/planet are given by $v \simeq \sqrt{2GM_*/q}$ 
and $L \simeq \sqrt{2 GM_* q}$ 
that are independent of initial semimajor axis $a$ of incoming bodies. 

\begin{figure}[htbp]
\plotone{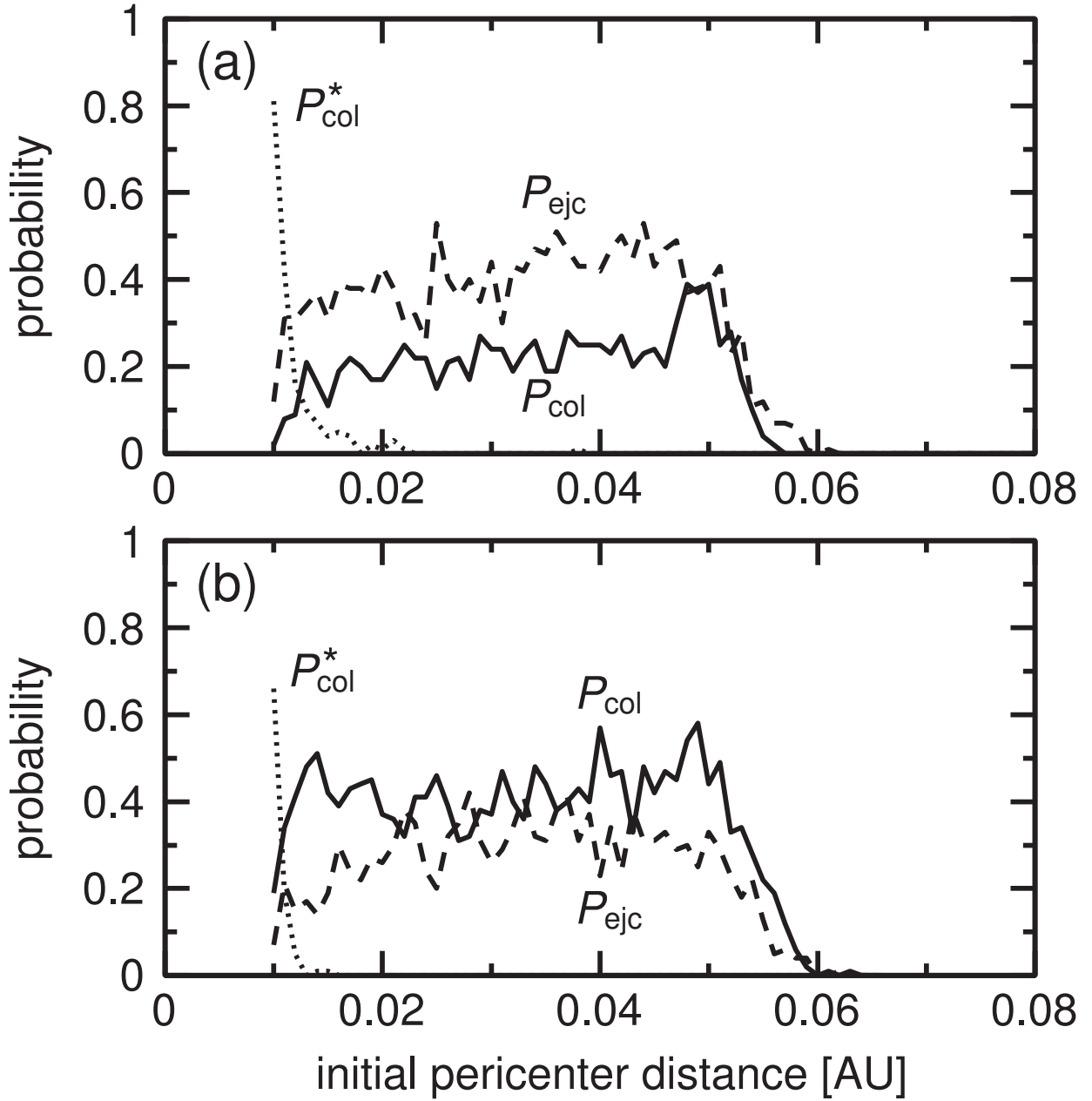}
\caption{
Probability of outcomes
of encounters between the inner giant planet of $0.5M_{\rm J}$
in a circular orbit at 0.05~AU and a planetesimal (test particle) 
(panel a) or another giant planet of $0.5M_{\rm J}$ (panel b) 
in a nearly parabolic orbit with initial semimajor axis $a = 1$~AU.
For each initial $q$, 100 cases with
random angular distributions are calculated until 100 Keplerian periods
at 1~AU.
Collision probability with the inner planet
($P_{\rm col}$), that with the parent star ($P_{\rm col}^*$),
and ejection probability ($P_{\rm ejc}$) are expressed by 
solid, dotted, and broken lines. 
Here, the physical size of planets are determined with
internal density 1 g\,cm$^{-3}$;
that of the parent star is set as 0.01~AU.
}
\label{fig:scat}
\end{figure}

Figure~\ref{fig:scat} shows the probabilities of the three outcomes of encounters 
between the inner planet and a planetesimal (panel a) or 
another giant planet (panel b). 
As shown in Fig.~\ref{fig:scat}, 
when $0.01 {\rm AU} < q < 0.06 {\rm AU}$, the incoming bodies 
closely approach the inner planet.
Although some fraction of them are ejected, 
the comparable fraction results in collision with the inner planet. 
In particular, $P_{\rm col} > P_{\rm ejc}$ in the planet-planet case (panel b), 
in which the ratio of geometrical cross section to Hill radius 
is larger than in the planet-planetesimal case (panel a).
If $q$ is smaller than or close to the parent star's physical 
radius (which is 0.01~AU in the calculations here), 
most of the incoming bodies hit the parent star 
rather than collide with the inner planet. 
If $q> 0.06$~AU, they do not closely encounter with the inner planet, so that
both collision and ejection probabilities almost vanish
($P_{\rm col}, P_{\rm ejc} \ll 1$).

The results shown in Fig.~\ref{fig:scat} demonstrate that 
efficient supply of heavy elements to the inner giant planet requires 
a very limited range (close to unity) of eccentricities of the incoming bodies; 
for example, $0.94 < e < 0.99$ for $a = 1$~AU, since $q = a(1-e)$.
In the case of chaotic scattering by outer giant planets,
the probability to acquire $q$ in such a narrow range would be
very small \citep*[e.g.,][]{Ford2001}.
On the other hand, if the scattering comes from secular perturbation,
$e$ is increased secularly while $a$ is kept constant.
Accordingly, $q$ decreases secularly, and 
the incoming bodies eventually collide with the inner planet.

For example, the following scenario is possible.
More than three giant planets in addition to the inner one are formed in nearly
circular orbits. 
They start orbit crossing on timescales
longer than their formation timescales \citep{RF96,WM96,LI97,MW02}. 
In many cases, one giant planet is ejected and the residual giant planets
remain in stable eccentric orbits after the orbit crossing
\citep{MW02}. 
In about 1/3 cases, $q$ of a giant planet 
decreases secularly to $\la 0.05$~AU due to the Kozai effect \citep{Kozai62}
during the orbit crossing \citep{nagasawa}.
This scenario requires formation of at least four giant planets.

\subsection{Origin of the Small Amount of Hydrogen and Helium 
\label{sec:origin_of_small_envelope}}

\subsubsection{Limited disk gas supply \label{sec:limited_gas_supply}}

Gas accretion onto a core can be truncated by gap opening
or global depletion of disk gas.
A gap may be formed, if both the viscous and thermal
conditions are satisfied \citep[e.g.,][]{LP85, LP93,CMM06}.  
The former is that the torque exerted by the planet overwhelms
viscous torque, and given by \citep{IL04a},
\begin{equation}
M_{\rm p} > M_{\rm p,vis} = 30 \left(\frac{\alpha}{10^{-3}}\right)
            \left(\frac{a}{1{\rm AU}}\right)^{1/2} 
            \left(\frac{M_*}{{\rm M}_{\odot}}\right) \mearth,
\label{eq:vis_condition}
\end{equation}
where $M_{\rm p}$ is the planet mass
and $\alpha$ is the parameter for the $\alpha$-prescription
for disk viscosity \citep{alpha}.  
The latter condition is that the Hill radius becomes
larger than disk scale height, and given by\footnote{
Here we assume the Hill radius ($r\sub{H,c}$) is equal 
to 1.5 times disk scale height ($h$), 
roughly taking into account late gas accretion with a reduced rate 
after $r\sub{H, c} = h$, following \citet{IL04a}.
}
 \citep{IL04a}
\begin{equation}
M_{\rm p} > M_{\rm p,th} = 400 
            \left(\frac{a}{1{\rm AU}}\right)^{3/4} 
            \left(\frac{M_*}{{\rm M}_{\odot}}\right) \mearth,
\label{eq:therm_condition}
\end{equation}
assuming an optically thin disk in which $T = 280(a/1{\rm AU})^{-1/2}$~K.
Hence, the actual truncation condition may be
the thermal condition and it is  very unlikely that gas accretion 
is truncated by the gap opening at $M_{\rm p} \sim 110 \mearth$ 
(which corresponds to HD149026b) far from the parent star, 
as long as a sufficient amount of disk gas remains. 
Furthermore, since the gap is replenished by viscous diffusion, 
gas accretion may not completely be truncated  
by the gap opening \citep*{D'Angelo03}.

If a sufficient amount of disk gas remains, 
one way to truncate gas accretion at relatively small planetary mass
is that the planet migrates to the vicinity of its parent star
before the planet accretes a large amount of gas.
Both $M_{\rm p,vis}$ and $M_{\rm p,th}$ are small in the 
inner regions (Eqs.~[\ref{eq:vis_condition}] and [\ref{eq:therm_condition}]).
However, 
this is unlikely, 
even if type II migration occurs when $M_{\rm p} \sim 30 \mearth$. 
The timescale for the KH gas accretion $\la 3 \times 10^4$ years 
for a core of $\ga 30 \mearth$ (Fig.~\ref{fig:KH}), 
while the timescale for the migration is much longer ($\sim 10^6$--$10^7$ years). 
Furthermore, it is not clear if the gap opening can 
stop gas accretion completely.

If disk gas is globally depleted 
when $M_{\rm c}$ reaches $M_{\rm c,crit}$, 
gas accretion onto the core can be limited 
by viscous diffusion of disk gas, 
not by the Kelvin-Helmholtz contraction of the envelope
\citep[e.g.,][]{GuillotHueso06}. 
That could be possible to account for the small amount of H/He. 
However, whether such a small amount of disk gas can 
bring the planet to the vicinity of the parent star should be examined.
If it does not work, different migration mechanism such as gravitational
scattering during orbital crossing of giant planets is required.

\subsubsection{Loss of envelope gas \label{sec:envelope_loss}}

Another way to explain the small amount of the H/He gas of HD149026b 
is loss of the envelope gas. 
There are three possibilities for loss of the envelope gas; 
photoevaporation driven by incident UV flux from the parent star, 
the Roche lobe overflow, and 
impact erosion by a collision with another gas giant planet.

%\noindent
%{\bf a) Photoevaporation}

Photoevaporation process for gas-rich planets is normally faster for
smaller planetary masses, because less massive planets are more
expanded and their envelope gas is more loosely bound
\citep{Baraffe06}.  This means the envelope quickly disappears 
once the evaporation occurs, so that it should be relatively rare to
observe a planet at a stage when a relatively small amount (30--60
$\mearth$) of envelope gas remains.
However, this possibility is not excluded at present, because $Z$-rich
planets such as HD149026b can be more compact and their envelope gas
is not necessarily more loosely bound \citep[e.g.,][]{BACB06}.  
%In that case, H/He components
%may have to selectively evaporate from the planet to form a
%$Z$-enriched envelope.  But the evaporation may be hydrodynamic.  It
%is not clear if such selective evaporation can occur in the rapid
%process.
%
%Although observed \citep{VidalMadjar04}, the magnitude of the
%evaporation process is very much in question as uncertainties of up to
%3 orders of magnitude exist in models of the process
%\citep[e.g.,][]{Baraffe06}. 

%\noindent
%{\bf b) Roche lobe overflow}

Envelope gas can also be lost by the Roche lobe overflow. 
When the Roche lobe is filled with the envelope gas, 
the gas overflows through the inner Lagrangian point to the inner disk. 
The Roche lobe overflow takes place 
if a planet is very close to its parent star 
because of inflation by stellar tidal heating as well as 
reduction of the Roche lobe \citep[e.g.,][]{Trilling98}, 
and subsequent tidal inflation instability of the envelope 
\citep*[e.g.,][]{Gu2003}.  
The spilled gas from the Roche lobe first goes to the inner disk 
and is eventually going to fall on the parent star 
by the gravitational interaction with the planet. 
At the same time, the planet gains angular momentum as the counteraction 
and migrates outward. 
Because of the increase in the orbital radius, 
the evaporation process slows down. 
Thus a state of a 30--60~$\mearth$ envelope could be possible. 
This possibility is, however, in question 
because it is unclear if sufficient angular momentum is lost.

%\noindent
%{\bf c) Giant impacts}

\citet{Sato05} suggested that collision between two giant planets 
could account for the internal structure of HD149026b.
If cores with individual mass of $\sim 30 \mearth$ are merged 
while significant amounts of their envelopes evaporate,
a planet similar to HD149026b is formed.
Here we examine if significant loss of the envelope gas occurs 
through SPH simulations of a collision between two giant planets.

High impact velocity is expected. 
Suppose that one giant planet orbits at 0.03--0.05~AU and
another giant planet in a highly eccentric orbit is sent from an outer region 
(for the mechanism to send planets, see section 
\ref{sec:late_Z_supply}).
Since the impactor has high orbital
eccentricity ($e \sim 1$),
collision velocity is similar to 
local Keplerian velocity at 0.03--0.05~AU around a $1.3M_{\odot}$ star,
150--190 km/s. 
This velocity is 2--3 times as high as 
the two-body surface escape velocity. 
Since specific angular momentum of the impactor is given by
$L = \sqrt{G M_* a(1-e^2)} \simeq \sqrt{2 G M_* q}$, 
the impactor and the target have similar $L$.
Hence, the semimajor axis of the merged planet would be similar to
that of the target planet unless envelope evaporation changes
specific angular momentum significantly.
 
Previous SPH simulations of a collision between Mars-size or Earth-size 
rocky planets \citep[e.g.,][]{Canup04, AA04} show
almost no loss of material occurs by a low-velocity impact ($v_{\rm imp}$) of 
$1.0$--$1.5$ times as high as the two-body surface escape velocity 
($v_{\rm esc}$) defined by 
$v_{\rm esc}^2 = 2G(M_{\rm t}+M_{\rm i})/(R_{\rm t}+R_{\rm i})$, 
where $M$ and $R$ are the planetary mass and radius and 
subscriptions t and i represent the target and the impactor.
Collisions with higher impact velocity lose larger amounts of material; 
for example, 40\% of the total mass is lost when $v_{\rm imp} \sim 2.5 v_{\rm esc}$ 
in their simulations. 
For gas giant planets, we will have qualitatively similar results, 
but can expect more significant loss of envelope gas 
because the envelope is less tightly bound compared to rocky planets. 

In order to simulate the collision between two gas giant planets, 
we use a SPH method with the modified spline kernel function 
\citep{ML85} and artificial viscosity \citep{MG83}. 
Our SPH simulation code was checked to reproduce the results of \citet{AA04} 
for Mars-size planets using the Tillotson EOS. 
In this paper, 
we assume that the gas giant planets are composed only of ideal gas with 
the adiabatic exponent of 2; 
this value reproduces the present structure of Jupiter. 
We consider two hydrostatically equilibrated planets 
(each being composed of about 30,000 SPH particles) with Jupiter's mass and radius
for initial conditions. 
We perform the simulations of head-on collisions with various impact 
velocities (1.0, 1.5, 2.0, 2.25, 2.5, and 3.0$v_{\rm esc}$). 
In our calculations, $v_{\rm esc}$ is 60 km/s. 
We calculate the gravitational force for each SPH particle using 
the special-purpose computer for gravitational N-body systems, GRAPE-6.

\begin{figure}[htbp]
\plotone{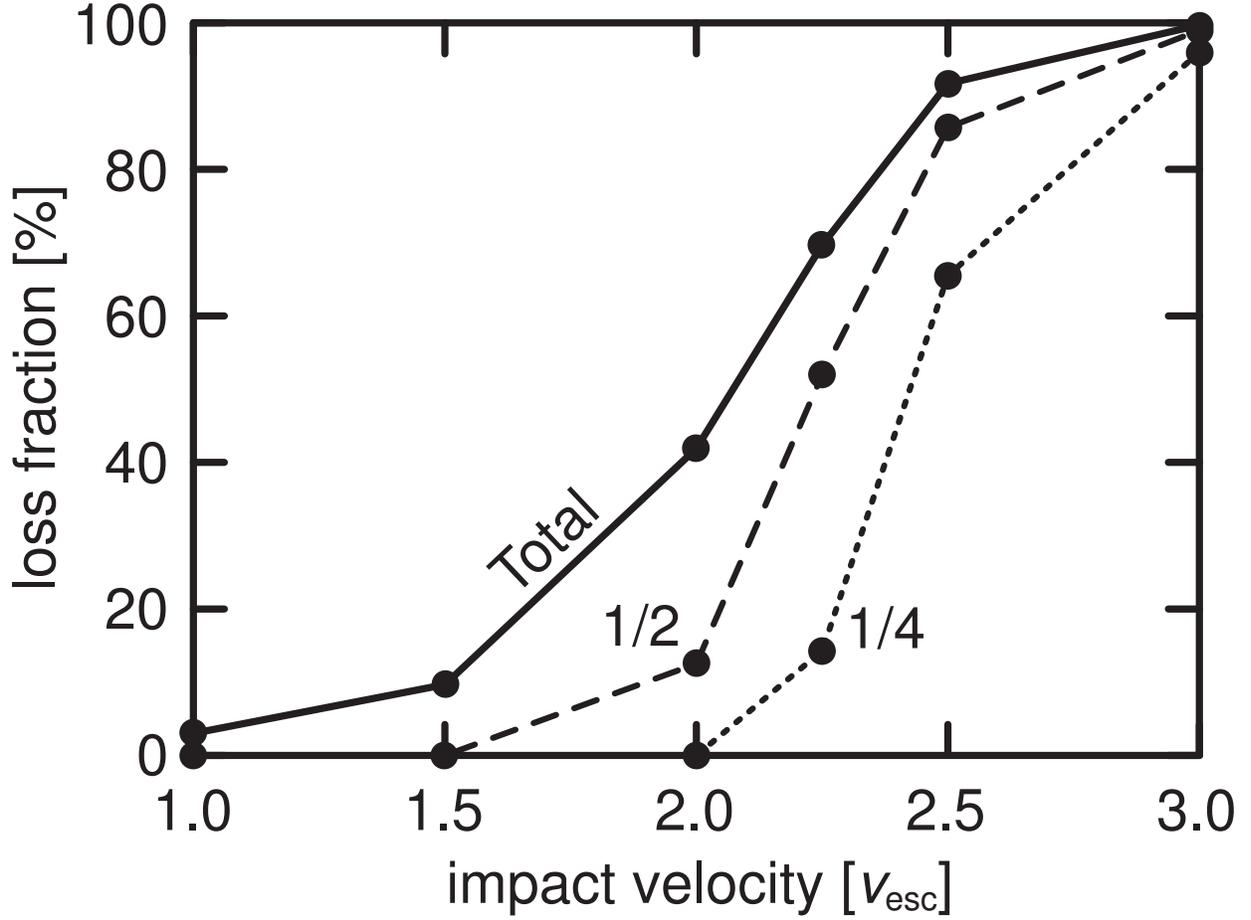}
\caption{
The loss fraction of mass after a head-on collision 
between two Jupiter-mass planets composed of ideal gas 
with the adiabatic exponent of 2. 
The loss fraction of mass 
initially located inside 1/2 (dashed line) 
and 1/4 (dotted line) of the initial planetary radius
are plotted to compare with total loss fraction (solid line). 
The loss fraction is defined as the mass fraction of 
the SPH particles whose velocity exceeds the local escape velocity 
and distance from the center of mass frame exceeds 5 times 
of the initial planetary radius.
}
\label{fig8}
\end{figure}

Figure \ref{fig8} shows the mass fraction lost 
by collisions with various impact velocities. 
Higher impact velocity results in higher loss fraction. 
Almost no escape occurs for a low-velocity impact 
($v_{\rm imp} \sim v_{\rm esc}$).
%which is consistent with \citet{Canup04} and \citet{AA04}. 
When the impact velocity is higher than $2.5 v_{\rm esc}$, 
almost all SPH particles ($\sim$ 90\%) are lost. 
%These results are somewhat different from ones 
%by \citet{AA04}, who considered collision 
%between two Mars-sized planets, and used the Tillotson EOS. 
%Since we have reproduced their results with the same EOS 
%on almost the same conditions, 
%the difference in results between ours and \citet{AA04} 
%might be due to difference in EOS. 

Although we are unable to know exactly the loss fraction of the core 
with our current simulation, 
we can estimated it approximately by tracking 
the motion of the SPH particles which are initially located 
in inner part of the planet. 
Figure \ref{fig8} also shows the loss fractions of the inner parts that are 
initially located inside 1/2 and 1/4 of the planetary radius. 
The inner part of the planet tends to remain in the merged planet. 
For example, in the case of $2.0 v_{\rm esc}$, 
about 50\% of materials is totally lost, but almost no escape occurs 
for the material inside of 1/4 initial radius.
A head-on collision of two gas giant planets with 
$v_{\rm imp} \sim 2$--$2.5 v_{\rm esc}$ possibly results 
in the considerable depletion of envelope and merging of core.

The strongest point of this model for envelope loss is 
that we need no additional process to supply heavy elements to the planet.
The collision increases $M_Z$ up to $\sim 60 \mearth$
as well as decreases $M_{\rm H\,He}$;
as we showed in section \ref{sec:subcritical}, 
it is possible to form cores of $\sim 30{\rm M}_{\oplus}$
by ordinary planetesimal accretion in relatively massive disks,
in particular, metal-rich disks.
However, the numerical simulations performed here is still preliminary. 
Since this mechanism may be promising, 
we should perform detailed simulations in the future.

\subsection{Other Possibilities \label{sec:discussion}}

Another possibility of formation of a metal-rich giant planet is 
the formation in a disk with originally high dust/gas ratio. 
In general, dust grains migrate inward in a gas disk to change
the local dust/gas ratio \citep[e.g.,][]{Stepinski97}.
If the dust/gas ratio is enhanced in planet-forming regions
at $\la 10$~AU, the metal-enriched envelope could be formed without 
any planetesimal/planet supply.
The high dust/gas ratio also makes possible planetesimal formation from
dust through self-gravitational instability against Kelvin-Helmholtz
instability \citep{Youdin02} and avoids 
type-I migration of terrestrial planets \citep*{Kominami05}.
However, it is not clear if the dust/gas ratio
can become large enough to account for $M_Z$ of HD149026b.

In principle, it is possible for a $70{\rm M}_{\oplus}$ core 
to form at $\sim 0.05$~AU
through accretion of many Earth-size to Uranian-size cores that 
migrate close to the parent star.
In the vicinity of the parent star, 
gas accretion should be truncated at a relatively small mass 
(see eq.~[\ref{eq:therm_condition}]).
We may need to carry out N-body simulations of the migrating
protoplanets in order to examine whether they can pass through
resonant trapping (or shepherding) to merge or not.
However, we should keep in mind that incorporation of type-I migration 
without any conditions causes severe inconsistency with
observed data of extrasolar planets and our Solar System.
If we rely on type-I migration model, we need to clarify
the condition for the occurrence of type-I migration at the same time.

\section{SUMMARY AND CONCLUSIONS \label{sec:summary}}

The high density of the close-in extrasolar giant planet HD149026b 
recently discovered by \citet{Sato05} challenges theories of planet formation. 
In this paper, we have attempted to derive robust constraints on the
planet's composition and infer possible routes to explain its
formation. 

We have first simulated the evolution of HD149026b more extensively 
than previous workers \citep{Sato05,Fortney06} 
and confirmed that the planet contains a substantial amount of heavy elements. 
Preferred values of the total mass of heavy elements are 50--80~$\mearth$ 
(section~\ref{sec:standard_model}), 
which is consistent with the previous calculations.
We showed that the results are unchanged for heavy elements located in the 
central core, or distributed inside the envelope, provided they remain
deeper than the external radiative zone. In the event of a significant
enrichment of the outer layers, slightly higher values of heavy
elements content are possible (section~\ref{sec:core_or_envelope}). 
In order to derive minimum values of the mass of heavy elements, we
have explored the possibility that the planet was stored in a relatively
cold environment for some time before migrating near to the
planet. This strict minimum is $\sim 35 \mearth$, but is
regarded as unlikely because it requires a late migration and no
reheating of the planet by  tidal circulization 
(section~\ref{sec:cold_storage_hypothesis}). 

We have then investigated the possibility of subcritical core accretion 
as envisioned for Uranus and Neptune 
to account for the small envelope mass as well as the large core mass. 
In principle a large core of 50--80~$\mearth$ can be formed 
by subcritical core accretion. 
However we have found it very unlikely for at least two reasons: 
(i) A subcritical core accretion results in a ratio of the core mass
to the total mass above $\sim 0.7$ (section~\ref{sec:critical_mass}), 
whereas our evolution calculations showed such a high ratio to be possible 
in a very limited range of parameters (see Table~\ref{tab:constraints});
(ii) The subcritical formation of a 50--80 $\mearth$ core requires 
an extremely massive or metal-rich disk with dust surface density 
30-50 times the values obtained for the minimum mass solar nebula
(section~\ref{sec:subcritical}). 
A reasonably massive and/or metal-rich disk can form 
cores of at most $\sim 30 \mearth$ far from the parent star.  
Those facts require us to consider (i) the migration of the planet, 
(ii) the supply of heavy elements to the planet during or after the gas
accretion phase,  
and (iii) a limited supply of disk gas or loss of the envelope gas 
to account for the properties of HD149026b. 

In section~\ref{sec:Z_supply}, we have discussed how the heavy elements can
be delivered to the planet during or after the gas accretion phase
according to these scenarios. 
An efficient delivery during the gas accretion phase needs 
to be re-investigated in much more details, 
because the shepherding tends 
to prevent the planet from accreting planetesimals 
(section~\ref{sec:concurrent_Z_supply}). 
On the other hand, scattering of planetesimals/planets by
one or several outer giant planets was shown to lead to an efficient
accretion by a close-in giant planet, and is a promising explanation
for the formation of metal-rich planets 
(section~\ref{sec:late_Z_supply}). 

The relatively small amount of hydrogen and helium present in
HD149026b is also to be explained by the relatively slow
formation of the planet in a low-density environment
(section~\ref{sec:limited_gas_supply}). Another possibility is related to
erosion of the envelope during giant impacts. 
A reasonable impact velocity of 2--2.5 times the two-body surface
escape velocity was found to result in a substantial loss of the envelope gas, 
while solid cores are probably merged (section~\ref{sec:envelope_loss}). 

In summary, we can propose at least two scenarios for the origin of HD149026b. 
\begin{itemize}
\item A giant planet with a core of $\sim 30 \mearth$ forms 
      far from its parent star in a relatively massive and/or metal-rich disk.  
      Then it moves to the vicinity of the parent star 
      through type-II migration \citep{L96}. 
      In the outer regions additional more than three giant planets form, 
      which may be likely in massive/metal-rich disks \citep{IL04a, IL04b}. 
      They starts orbit crossing and the innermost one is temporally detached 
      from the outer two. The perturbation from the outer two secularly 
      pumps up the eccentricity of the innermost one \citep{nagasawa}. 
      The associated secular decrease of its pericenter distance results in 
      collision with the close-in planet. 
      The impactor also has a core of $\sim 30 \mearth$. 
      Their solid cores are merged to form a core of $\sim 60 \mearth$, 
      while envelope gas is severely eroded to $M_{\rm H He} \sim 50 \mearth$. 
      The orbital eccentricity of the merged body is damped by tidal
      interaction with the parent star. 
\item A $\sim 30 \mearth$ core forms in an evaporating disk 
      that was originally massive and/or metal rich. The core thus becomes
      supercritical, but the accretion of the envelope is limited by
      viscous diffusion in an evolved, extended gaseous disk 
      \citep[see][]{GuillotHueso06}. 
      The planet subsequently accretes heavy
      elements through the delivery of planetesimals and/or planets,
      either because of migration due to interactions with the
      gas disk or to secular perturbations with a massive outer
      planet. 
\end{itemize}

Above two scenarios are different from each other 
in that the former leaves at least one outer giant planet 
while the latter does not necessarily need any outer giant planets. 
Long-term radial velocity observations of HD149026b are thus needed to
determine the presence of other planets in the system.

% Acknowledgement
We thank the anonymous referee for fruitful comments. 
This research was supported by 
Ministry of Education, Culture, Sports, Science and Technology of Japan, 
Grant-in-Aid for Scientific Research on Priority Areas, 
``Development of Extra-solar Planetary Science''.

\clearpage

\clearpage

\clearpage

\clearpage

\clearpage

\clearpage

\clearpage

\clearpage

\end{document}